\documentclass[utf8]{frontiersSCNS} 

\usepackage{url,hyperref,microtype,subcaption}
\usepackage[onehalfspacing]{setspace}

\def\firstAuthorLast{N\'u\~nez Fern\'andez {et~al.}} 
\def\Authors{Y. N\'u\~nez Fern\'andez\,$^{1,*}$ and K. Hallberg\,$^{1}$}
\def\Address{$^{1}$ Centro At\'omico Bariloche and Instituto Balseiro, CNEA, CONICET, 8400 Bariloche, Argentina }
\def\corrAuthor{Y. N\'u\~nez Fern\'andez}

\def\corrEmail{yurielnf@gmail.com}

\begin{document}
\onecolumn
\firstpage{1}

\title[Efficient multi-site and multi-orbital DMFT]{Solving the multi-site and multi-orbital
Dynamical Mean Field Theory using Density Matrix Renormalization} 

\author[\firstAuthorLast ]{\Authors} 
\address{} 
\correspondance{} 

\extraAuth{}

\maketitle

\begin{abstract}
We implement an efficient numerical method to calculate response functions of complex impurities based on the Density Matrix Renormalization Group (DMRG) and use it as the impurity-solver of the Dynamical Mean Field Theory (DMFT). This method uses the correction vector to obtain precise Green's functions on the real frequency axis at zero temperature. By using a
self-consistent bath configuration with very low entanglement, we take full advantage of the DMRG to calculate dynamical response functions paving the way to treat large effective impurities such as those corresponding to multi-orbital interacting models and multi-site or multi-momenta clusters. This method leads to reliable calculations of non-local self energies at arbitrary dopings and interactions and at any energy scale.

\end{abstract}

\section{Introduction}
Among the most intriguing problems in physics is the behaviour of strongly correlated materials which present emergent behavior such as high temperature superconductivity, ferroelectricity, magnetism and metal-insulator transitions. These systems have triggered a great deal of research and are still far from being understood. However, a complete theoretical understanding is still lacking due to the presence of strongly interacting local orbitals in these materials. Methods to calculate electronic structure of weakly correlated materials, such as the Density Functional Theory (DFT)  \citep{hohenberg} which use the local density approximation (LDA)  \citep{jones} and other generalizations, are unable to describe accurately the strong electronic correlation case. Non-perturbative numerical methods are, thus, the only reliable approach.

To include correlations, the Dynamical Mean Field Theory (DMFT) was developed more than twenty years ago. Together with its sucessive improvements  \citep{pt,review,savrasov2001,maier,hettler,senechal}, these methods have led to more reliable results.  The combination of the DMFT with LDA has allowed for band structure calculations of a large variety of correlated materials (for reviews see Refs.  \citep{imada,Held}), where the DMFT accounts mainly for local interactions  \citep{anisimov,katsnelson}. A recent alternative proposal, the Density Matrix Embedding Theory, DMET, was developed, which relies on the embedding of the wave functions of a local cluster fragment (instead of the local Green functions) in a self-consistent finite environment  \citep{Chan1,Bulik}.

The DMFT requires the calculation of an interacting quantum impurity for which the fermionic environment has to be determined self-consistently until convergence of the local Green functions and the local self-energies is reached. Therefore, the success and scope of the DMFT will depend on the existence of accurate methods to solve correlated and complex quantum impurities. This approach is exact for the infinitely coordinated system (infinite dimensions), the non-interacting model and in the atomic limit.  

Several quantum impurity solvers have been proposed since the development of the DMFT,  among which we can mention the iterative perturbation theory (IPT)  \citep{GeorgesKotliar,RozenbergKotliar}, exact diagonalization (ED)  \citep{Caffarel,gunnarsson}, the Hirsch-Fye quantum Monte Carlo (HFQMC)  \citep{hf}, the continuous time quantum Monte Carlo (CTQMC)  \citep{ctqmc1,ctqmc2,ctqmc3,CTQMC,ryo}, non-crossing approximations (NCA)  \citep{NCA}, the numerical renormalization group (NRG)  \citep{wrg,bulla,weichselbaum,stadler}, the rotationally invariant slave-boson mean-field theory (RISB)  \citep{Lechermann2007,PhysRevB.80.115120,Ferrero2009} and quantum chemistry-based techniques  \citep{ChanCI}.
Although these methods allow for the calculation of relevant properties such as the metal-insulator transition and other low-lying energy properties, they present some problems. Among them, one can mention the sign problem and the difficulty in reaching low temperatures in the QMC-based algorithms, the failure of the NCA in obtaining a reliable solution for the metallic state, the limitation to few lattice sites, far from the thermodynamic limit of the ED and the reduced high-energy resolution of the NRG technique. 

To overcome some of these difficulties an impurity solver based on the Densit Matrix Renormalization Group (DMRG) technique  \citep{white1, book, scholl, karen1, nishimoto} was proposed  \citep{garciadmft,garciadmft2,yuriel,karski}. Subsequent improvements to this were introduced, such as those using the time evolution DMRG algorithm  \citep{realtime,fork}, dynamical calculations using the Kernel Polynomial Method (Chebyshev expansion for Green functions)  \citep{weisse,holzner,wolf,ganahl2} and the application to non-equilibrium DMFT using MPS  \citep{wolf2}. It was recently realized that converging the DMFT loop on the the imaginary-frequency axis rather than the real-frequency axis reduces computational costs by orders of magnitude because the bath can be represented in a controlled way with far fewer bath sites and, crucially, the imaginary-time evolution does not create entanglement. The imaginary time setup can therefore treat much more sophisticated model Hamiltonians, opening the possibility of studying more complicated and realistic models and performing cluster dynamical mean field calculations for multiorbital situations. The price to be paid, however, is a reduced resolution on the real-frequency axis  \citep{Go}. 
 
In spite of these developments, several difficulties still remain which hinder the calculation of reliable spectral densities for complex multi-band and multi-orbital correlated systems  \citep{EPLHallberg}. In this paper we present a novel technique based on the DMRG which includes important improvements and complements previous methods. It is based on an efficient selection of the relevant states due to low entanglement bath configurations and on the targetting of the correction vector for small real energy windows. This method, thus, provides detailed spectral functions for complex Hamiltonians at zero temperature and for any doping and correlations. In the following sections we describe the method and show some applications and potential uses.



\section{General formulation}

In order to present a unified treatment of multi-site (or cluster)
and multi-orbital Hamiltonians on the lattice, we start by interpreting
the lattice as a superlattice such that: 
\begin{enumerate}
\item The interaction $\hat{V}$ is local and completely contained in
the unit cell: $\hat{V}=\sum_{i}\hat{V}_{i}^ {}$, where $i$ is the
cell index. 
\item The non-interacting Hamiltonian $\hat{H}^{0}$ is characterized by
its local Green function matrix $G_{0}(\omega\,\mathbf{1}-T)$; being
$T=\left(t_{IJ}\right)$ the coefficients of the local part $\hat{h}_{i}^{0}$
of $\hat{H}^{0}$: $\hat{h}_{i}^{0}=\sum_{IJ}t_{IJ}c_{iI\sigma}^{\dagger}c_{iJ\sigma}$,
where $c_{iI\sigma}^{\dagger}$ creates an electron in cell $i$ and
local ``orbital'' $I=1,2,..,N_{c}$ with spin $\sigma=\uparrow,\downarrow$.
\end{enumerate}
These two points completely define our problem through the parameters 
$\hat{V}_{i}$, $G_{0}$, $T$. Notice that $G_{0}$ and $T$ are typically
well known one-particle quantities for a given lattice problem.

The key idea of the DMFT is to neglect the self-energy between different
cells $i$ and $j$ in the lattice, that is, to consider only the
local self-energy: $\Sigma_{ij}(\omega)\approx\Sigma(\omega)\delta_{ij}$.
In this way, we are neglecting spatial correlations up to a certain degree while a good
treatment of the local dynamical correlations is made. The relevant point
is that the problem becomes tractable, as we will see in the following.
Note that $G_{0}$, $T$, and $\Sigma$ are $N_{c}\times N_{c}$
matrices for the spin-symmetric solution, and $2N_{c}\times2N_{c}$
matrices in the general case. Spatial correlations or the momentum dependence of $\Sigma$
can be obtain by periodization \citep{period}.

\begin{figure}
\begin{centering}
\includegraphics[width=4cm]{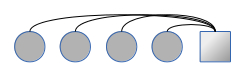}
\par\end{centering}

\begin{centering}
\includegraphics[width=8cm]{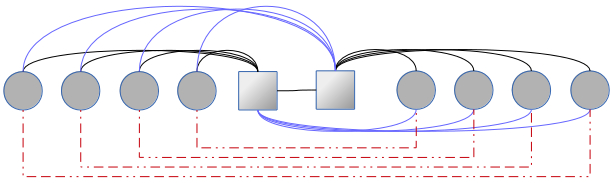}
\par\end{centering}

\begin{centering}
\includegraphics[width=8cm]{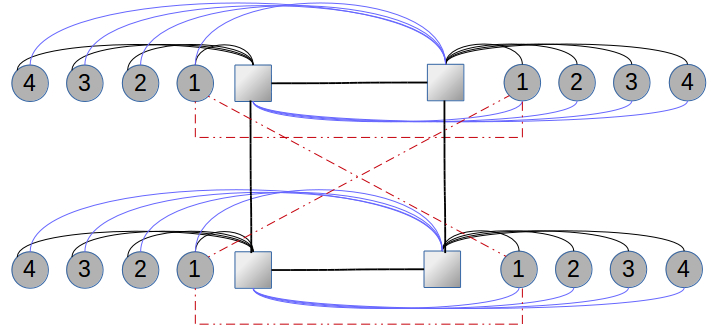}
\par\end{centering}

\caption{\label{fig:problem} Graphic representation of Hamiltonian Eq. (2) corresponding to the impurity problem for the one, two
and four-site cellular DMFT. The circles (squares) represent the non-interacting
(interacting/impurity) sites. The red lines correspond to the $\lambda_{q}^{IJ}$ parameters between bath sites $q$ related to impurities $I$ and $J$ (they are the only hybridization between the baths related to different impuritues). The blue lines are the $\upsilon_{q}^{IJ}$ with $I\ne J$ while the black lines are the $\upsilon_{q}^{II}$. In the bottom scheme we omit some obvious connections for clarity. 
}
\end{figure}

The local Green function is now given by \citep{dmft25}
\begin{equation}
G(\omega)=G_{0}\left(\omega\,\mathbf{1}-T-\Sigma(\omega)\right)
\end{equation}
which defines the \emph{self-consistency condition} for the $N_{c}\times N_{c}$
matrices $G$ and $\Sigma$. The lattice problem can now be mapped
onto an auxiliar impurity problem that has the same local magnitudes
$G(\omega)$ and $\Sigma(\omega)$. This impurity problem should be
determined iteratively. The impurity Hamiltonian can be written: 
\begin{equation}
H_{imp}=\hat{h}_{0}^{0}+\hat{V}_{0}+H_{b}\mbox{,}
\end{equation}
where the non-interacting part $H_{b}$ represents the bath: 
\begin{equation}
H_{b}=\sum_{IJq\sigma}\lambda_{q}^{IJ}b_{Iq\sigma}^{\dagger}b_{Jq\sigma}+\sum_{IJq}\upsilon_{q}^{IJ}\left[b_{Iq\sigma}^{\dagger}c_{0J\sigma}+H.c.\right]\mbox{,}
\end{equation}
$b_{Iq\sigma}^{\dagger}$ corresponds to the creation operator for
the bath-site $q$, associated to the ``orbital'' $I$ and spin
$\sigma$ (see Fig. \ref{fig:problem}). 

The self-consistent iterations can be summarized as follows:
(i) Start with $\Sigma(\omega)=0$,
(ii) Calculate the Green's function: 
\begin{equation}
G(\omega)=G_{0}\left(\omega-T-\Sigma(\omega)\right)\mbox{,}
\end{equation}

(iii) Obtain the hybridization
\begin{equation}
\Gamma(\omega)=\omega\,\mathbf{1}-T-\Sigma(\omega)-\left[G(\omega)\right]^{-1}\mbox{,}
\end{equation}

(iv) Find a Hamiltonian representation $H_{imp}$ with hybridization
$\tilde{\Gamma}(\omega)$ to approximate $\Gamma(\omega)$. The hybridization
$\tilde{\Gamma}(\omega)$ is characterized by the parameters $\Upsilon_{q}=\left(\upsilon_{q}^{IJ}\right)$
and $\Lambda_{q}=\left(\lambda_{q}^{IJ}\right)$ of $H_{b}$ through:
\begin{equation}
\tilde{\Gamma}(\omega)=\sum_{q}\Upsilon_{q}\cdot\left[\omega\,\mathbf{1}-\Lambda_{q}\right]^{-1}\cdot\Upsilon_{q}\mbox{.}\label{eq:hyb}
\end{equation}

(v) Calculate the impurity Green's function matrix $G_{imp}(\omega)$
of the Hamiltonian $H_{imp}$ using DMRG.
(vi) Obtain the self-energy 
\begin{equation}
\Sigma(\omega)=\omega\,\mathbf{1}-T-\left[G_{imp}(\omega)\right]^{-1}-\tilde{\Gamma}(\omega)\mbox{.}
\end{equation}
Return to \textbf{(ii)} until convergence.
Step (iv) is a fitting problem for $\Upsilon_{q}$ and $\Lambda_{q}$, where
we can use the general symmetries of the hybridization function matrix.
If $\Gamma$ can be diagonalized using the same unitary rotation $R$
for all $\omega$ then we obtain (at most) $N_{c}$ independent fittings.
This can be seen from Eq. (\ref{eq:hyb}) after applying $R$: 
\begin{equation}
R^{\dagger}\cdot\tilde{\Gamma}(\omega)\cdot R=R^{\dagger}\cdot\left(\sum_{q}\Upsilon_{q}\cdot\left[\omega\,I-\Lambda_{q}\right]^{-1}\cdot\Upsilon_{q}\right)\cdot R\mbox{,}
\end{equation}

\begin{equation}
\tilde{\Gamma}^{D}(\omega)=\sum_{q}\Upsilon_{q}^{D}\cdot\left[\omega\,\mathbf{1}-\mathbf{\Lambda}_{q}^{D}\right]^{-1}\cdot\Upsilon_{q}^{D}\mbox{,}\label{eq:hybD}
\end{equation}
where the superscript $D$ is used to stress that these matrices are
diagonal, and $M^{D}=R^{\dagger}\cdot M\cdot R$ where $M$ is an $N_{c}\times N_{c}$
matrix. In this new basis (the so-called molecular-orbital basis),
we have to fit $\Gamma_{11}^{D}(\omega)$ using the expression (\ref{eq:hybD})
for $\tilde{\Gamma}_{11}^{D}(\omega)$ which depends on the parameters
$\left(\Upsilon_{q}^{D}\right)_{11}$ and $\left(\Lambda_{q}^{D}\right)_{11}$, 
and similarly for $\Gamma_{22}^{D}(\omega)$, etc. Once these
independent fittings are done, we bring the parameters back to our
original basis through $M=R\cdot M^{D}\cdot R^{\dagger}$.

In general, symmetries can be expoited for a better performance and stability. For example, at half-filling we could also have the electron-hole symmetry, giving
a conection between $G(-\omega)$ and $G(\omega)$, implying the same
structure for the hybridization $\Gamma(\omega)$.

The most resource-demanding part of the algorithm is carried out at step (v), where the
dynamics of a complex many-body problem (see Figure \ref{fig:problem})
is calculated. 
Here we use the correction-vector method together with the DMRG
essentially following  \citep{kuhnerwhite,ramasesha}, although other methods to calculate dynamical response functions withing the DMRG can also be used  \citep{dindmrg,jeckelmann}. The one-dimensional representation
of the problem (needed for a DMRG calculation) is shown in the
figure, where we are also duplicating the graph when considering spin degress of freedom (not shown for clarity).
In this configuration (star geometry), in spite of the high connectivity of the Hamiltonian, the DMRG shows a much better performance  \citep{imaginarytime,wolfx}  

The correction vector method is implemented in DMRG by targeting not only the
ground state $\left|E_{0}\right\rangle $ of the system but also the
correction vector $\left|CV_{r}\right\rangle $ associated to the applied operator at 
frequency $\omega_{r}$ (and its neighborhood). For example, to obtain the single-particle density of states (DOS), the correction vector reads:
\[
\left(\omega_{r}+\mathbf{i}\eta-H_{imp}-E_{0}\right)\left|CV_{r}\right\rangle =c_{0I\sigma}^{\dagger}\left|E_{0}\right\rangle ,
\]
where a Lorentzian broadering $\eta$ was introduced to deal with
the poles of a finite-length impurity model. In this way a suitable
renormalized representation of the operators is obtained to calculate
the properties of the excitations around $\omega_{r}$, particularly
the Green's function, for instance,  $G_{JI\sigma}^{>}(\omega)=\left\langle E_{0}\right|c_{0J\sigma}\left|CV_{r}\right\rangle$ and $G_{JI\sigma}(\omega)=G_{JI\sigma}^{>}(\omega)+G_{JI\sigma}^{<}(-\omega)$.
Here $\omega_{r}$ with $r=1,2,...,N_{\omega}$
is a grid covering the frequencies of interest,
typically $N_{\omega}=$40-50 and are treated independently. Thus
each DMFT iteration uses around 30 cores totalling less than three hours for all cases considered in this work, considering system sizes of up to 36 sites.

\section{One-site DMFT }

As we remark, only three parameters should be defined in order to
apply the DMFT algorithm: $\hat{V}_{i}$, $G_{0}$, $T$. We study
the paramagnetic solution of the DMFT in the square (and Bethe) lattice
using the following: 
\[
\hat{V_{i}}=Un_{i\uparrow}n_{i\downarrow}\mbox{,}
\]
\[
T=-\mu\mbox{,}
\]
\[
G_{0}(\omega)=\begin{cases}
\frac{1}{N}\sum_{\mathbf{k}}\left[\omega-\epsilon(\mathbf{k})\right]^{-1} & \mbox{ Square lattice}\\
2\left[\omega+\sqrt{\omega^{2}-1}\right]^{-1} & \mbox{ Bethe lattice}
\end{cases}
\]
where $\epsilon(k)=-2t\left(\cos k_{x}+\cos k_{y}\right)-4t'\cos k_{x}cosk_{y}$,
with $\mathbf{k}=(k_{x},k_{y})$ the Fourier space of the square lattice
with $N$ sites, $N\rightarrow\infty$, and $t$ ( $t'$) denotes
the (next-)nearest-neighbor hopping integral \citep{ImadaHTc}.

\section{Two-band Bethe lattice}

We consider the interaction:

\begin{equation}
\begin{array}{c}
\hat{V_{i}}=U\sum_{I}n_{iI\uparrow}n_{iI\downarrow}+\sum_{\sigma\sigma'}\left(U_{2}-J\delta_{\sigma\sigma'}\right)n_{i1\sigma}n_{i2\sigma'}-\\
-J\left(c_{i1\uparrow}^{\dagger}c_{i1\downarrow}c_{i2\downarrow}^{\dagger}c_{j2\uparrow}+c_{i1\uparrow}^{\dagger}c_{i1\downarrow}c_{i2\downarrow}^{\dagger}c_{i2\uparrow}\right)\\
-J\left(c_{i1\uparrow}^{\dagger}c_{i1\downarrow}^{\dagger}c_{i2\uparrow}c_{i2\downarrow}+c_{i2\uparrow}^{\dagger}c_{i2\downarrow}^{\dagger}c_{i1\uparrow}c_{i1\downarrow}\right)
\end{array}\label{eq:interaction-1}
\end{equation}
where $J>0$ is the Hund exchange, $U$ ($U_{2}$) is the intra (inter)-orbital
Coulomb repulsion, and $I=1,2$ are the orbitals. The on-site non-interacting
coefficients are 
\[
T=\left(\begin{array}{cc}
-\mu & t_{12}\\
t_{12} & -\mu
\end{array}\right)\mbox{,}
\]
and the local Green's function:

\begin{equation}
G_{0}(\omega)=2\left[\omega\,\mathbf{1}+\sqrt{\omega^{2}\,\mathbf{1}-4B^{2}}\right]^{-1}\label{eq:latticeGreen}
\end{equation}
where $B=\left(\begin{array}{cc}
t_{1} & 0\\
0 & t_{2}
\end{array}\right)$, and $t_{1}$, $t_{2}$ are the nearest-neighbor hoppings for each
orbital.

Concerning step (iv), if $t_{12}=0$ then all our $2\times2$
matrices are diagonal and we have only to calculate two Green's functions
and do two independent fittings, one for each orbital. On the other
hand, if $t_{1}=t_{2}$ but $t_{12}\ne0$ then we can introduce the
rotation $R=\frac{1}{\sqrt{2}}\left(\begin{array}{cc}
1 & 1\\
1 & -1
\end{array}\right)$ to diagonalize the hybridization and we do again only two independent
fittings. In the general case, a non-diagonal matrix fitting should
be done to obtain a bath representation of the given hybridization
$\Gamma(\omega)$, that is, to find the parameters $\Upsilon_{q}$
and $\Lambda_{q}$ which minimize $\sum_{\omega}\left\Vert \Gamma(\omega)-\tilde{\Gamma}(\omega)\right\Vert ^{2}$
using, for instance, the matrix norm $\left\Vert M\right\Vert ^{2}=Tr\left[M^{T}\cdot M\right]$.

\begin{figure}
\begin{centering}
\includegraphics[width=11cm]{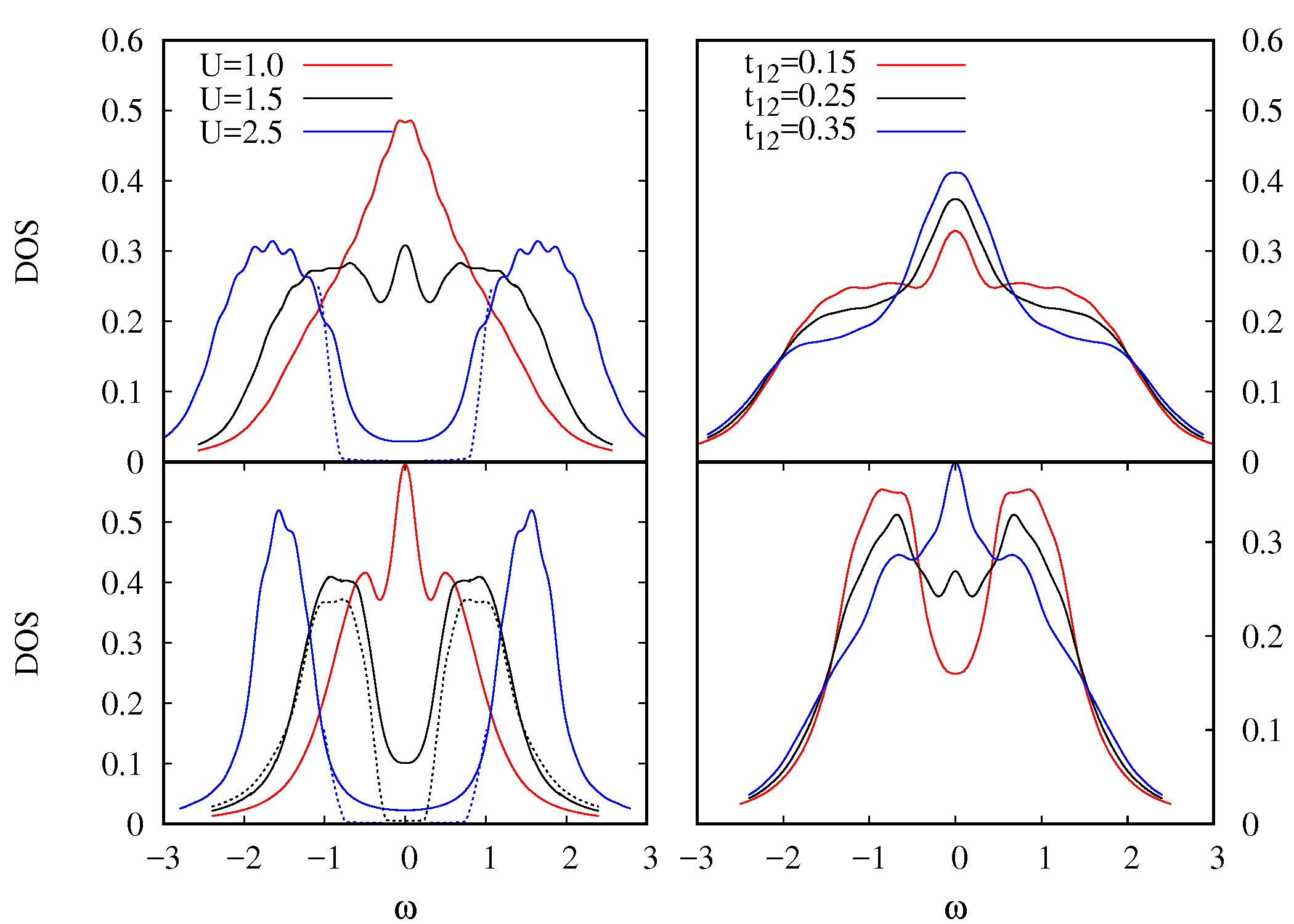}
\par\end{centering}
\caption{\label{fig2}DOS for the half-filled two-band Kanamori-Hubbard model on the Bethe lattice showing the orbital-selective Mott thansition for different bandwidths: $t_1=0.5$ (top panels) and $t_2=0.25$ (bottom panels). Left panel: varying $U$ for $t_{12}=0$. Right panel: varying $t_{12}$ for $U=1.5$. We consider the rotationally symmetric case $U_2=U-2J$ and $J=U/4$. Two different values of the broadening $\eta$ are depicted to emphasize the gapped region for the insulating phase.}
\end{figure}

In Fig. \ref{fig2} we present the results for this model where, by analyzing the DOS for the different bandwidth case, the orbital-selective Mott transition can be clearly observed for a finite Hund's coupling $J$.  This phase is robust for a certain range of interband hybridization. Previous calculations \citep{kogakawakami,kogados,demediciV} either resorted to approximate analytic continuation methods to obtain the DOS or used exact diagonalization for small baths. The method presented here leads to much more precise results and has the potentiality of treating even larger clusters or more orbitals. For example, it was crucial to find the in-gap holon-doublon quasiparticle peaks in the DOS of the asymmetric Hubbard model \citep{yurielhd}, which would have been hindered using QMC or NRG solvers or which would have lacked a proper finite size analysis had an ED method been used.

\section{Cellular DMFT on the square lattice}

We consider the same physical problem of section 3 on the square
lattice, but interpreted now in a superlattice of unit cell of size
$N_{c}=2\mbox{ (or \ensuremath{4})}$ corresponding to the two-site
(c2) or four-site (c4) cellular DMFT  \citep{imadacluster,kotliarcluster}. This case is illustrated in Fig. (1). The next-nearest-neighbor-hoppings $t'$ for the c4-DMFT connect the opposite vertices of the 4 impurity square depicted at the bottom of this figure.   
Our three parameters $\hat{V}_{i}$, $T$, $G_{0}$ are now:

\[
\hat{V_{i}}=U\sum_{I=1}^{N_{C}}n_{iI\uparrow}n_{iI\downarrow}\mbox{,}
\]

\[
T=\begin{cases}
\left(\begin{array}{cc}
-\mu & t\\
t & -\mu
\end{array}\right) & \mbox{ c2-DMFT}\\
\left(\begin{array}{cccc}
-\mu & t & t & t'\\
t & -\mu & t' & t\\
t & t' & -\mu & t\\
t' & t & t & -\mu
\end{array}\right) & \mbox{ c4-DMFT}
\end{cases}\mbox{,}
\]

\[
G_{0}(\omega)=\frac{N_{c}}{N}\sum_{\mathbf{\tilde{k}}}\left[\omega\,\mathbf{1}-\tilde{\epsilon}(\tilde{\mathbf{k}})\right]^{-1}\mbox{,}
\]
respectively. Here, $T$ is the non-interacting intracluster matrix and $\tilde{\epsilon}(\mathbf{\tilde{k}})$
is the intercluster hopping on the superlattice Fourier space $\mathbf{\tilde{k}}$,
which is connected to the one-site lattice through $\tilde{\epsilon}(\tilde{\mathbf{k}})_{IJ}=\frac{1}{N_{c}}\sum_{\mathbf{K}}\exp\left[i(\mathbf{K}+\tilde{\mathbf{k}})\cdot\mathbf{R}_{IJ}\right]\epsilon(\mathbf{K}+\tilde{\mathbf{k}})$
with $\mathbf{K}$ the intracluster Fourier-space vectors, see eq.
(23) of  \citep{maier}. 
\[
\tilde{\epsilon}(\mathbf{k})=\left\{ \begin{array}{c}
-t\left(\begin{array}{cc}
2\cos k_{y} & 1+\exp\left(2ik_{x}\right)\\
1+\exp\left(-2ik_{x}\right) & 2\cos k_{y}
\end{array}\right)-\\
t'\cos k_{y}\left(\begin{array}{cc}
0 & 1+\exp\left(2ik_{x}\right)\\
1+\exp\left(-2ik_{x}\right) & 0
\end{array}\right)
\end{array}\right.
\]
for the c2-DMFT and 
\[
\tilde{\epsilon}(\mathbf{k})=\left\{ \begin{array}{c}
-t\left(\begin{array}{cc}
0 & 1+\exp\left(2ik_{x}\right)\\
1+\exp\left(-2ik_{x}\right) & 0
\end{array}\right)\otimes\left(\begin{array}{cc}
1 & 0\\
0 & 1
\end{array}\right)-\\
t\left(\begin{array}{cc}
1 & 0\\
0 & 1
\end{array}\right)\otimes\left(\begin{array}{cc}
0 & 1+\exp\left(2ik_{y}\right)\\
1+\exp\left(-2ik_{y}\right) & 0
\end{array}\right)\\
-t'\left(\begin{array}{cc}
0 & 1+\exp\left(2ik_{x}\right)\\
1+\exp\left(-2ik_{x}\right) & 0
\end{array}\right)\otimes\\
\left(\begin{array}{cc}
0 & 1+\exp\left(2ik_{y}\right)\\
1+\exp\left(-2ik_{y}\right) & 0
\end{array}\right)\mbox{.}
\end{array}\right.
\]
for the c4-DMFT. Finally, the hybridization matrix $\Gamma$ has the
following form, see  \citep{imadacluster}: 
\[
\Gamma(\omega)=\begin{cases}
\left(\begin{array}{cc}
a & b\\
b & a
\end{array}\right) & \mbox{ c2-DMFT}\\
\left(\begin{array}{cccc}
a & b & b & c\\
b & a & c & b\\
b & c & a & b\\
c & b & b & a
\end{array}\right) & \mbox{ c4-DMFT}
\end{cases}
\]
which can be diagonalized using the corresponding unitary rotation
$R$, obtaining (at most) $N_{c}$ independent fittings.

\begin{figure}
\begin{centering}
\includegraphics[width=11cm]{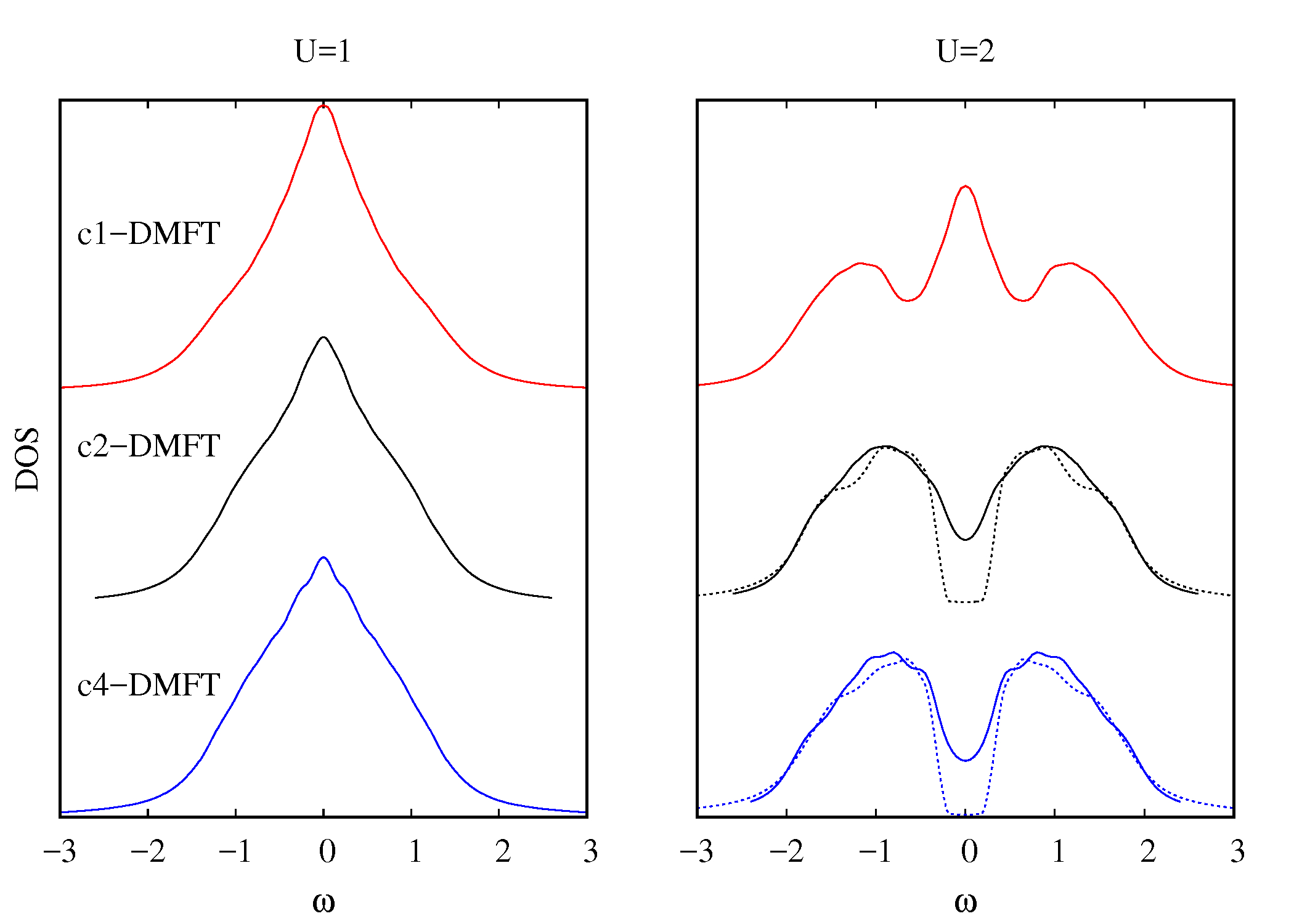}
\par\end{centering}
\caption{\label{fig3} Comparison of the DOS calculated using different cluster sizes for the Hubbard Hamiltonian on the square lattice for two values of $U$.  As in Fig. 2, results for two different values of the broadening $\eta$ are shown, emphasizing the gap for the insulating regime. The curves are shifted for clarity. }
\end{figure}

In Fig. \ref{fig3} we show the DOS for the Hubbard Hamiltonian on the square lattice with nearest ($t=0.25$) and next nearest-neighbor hopping ($t'=0$) for two values of $U$. Larger clusters lead to a smaller critical $U$'s \citep{imada}.

\begin{figure}
\begin{centering}
\includegraphics[width=11cm]{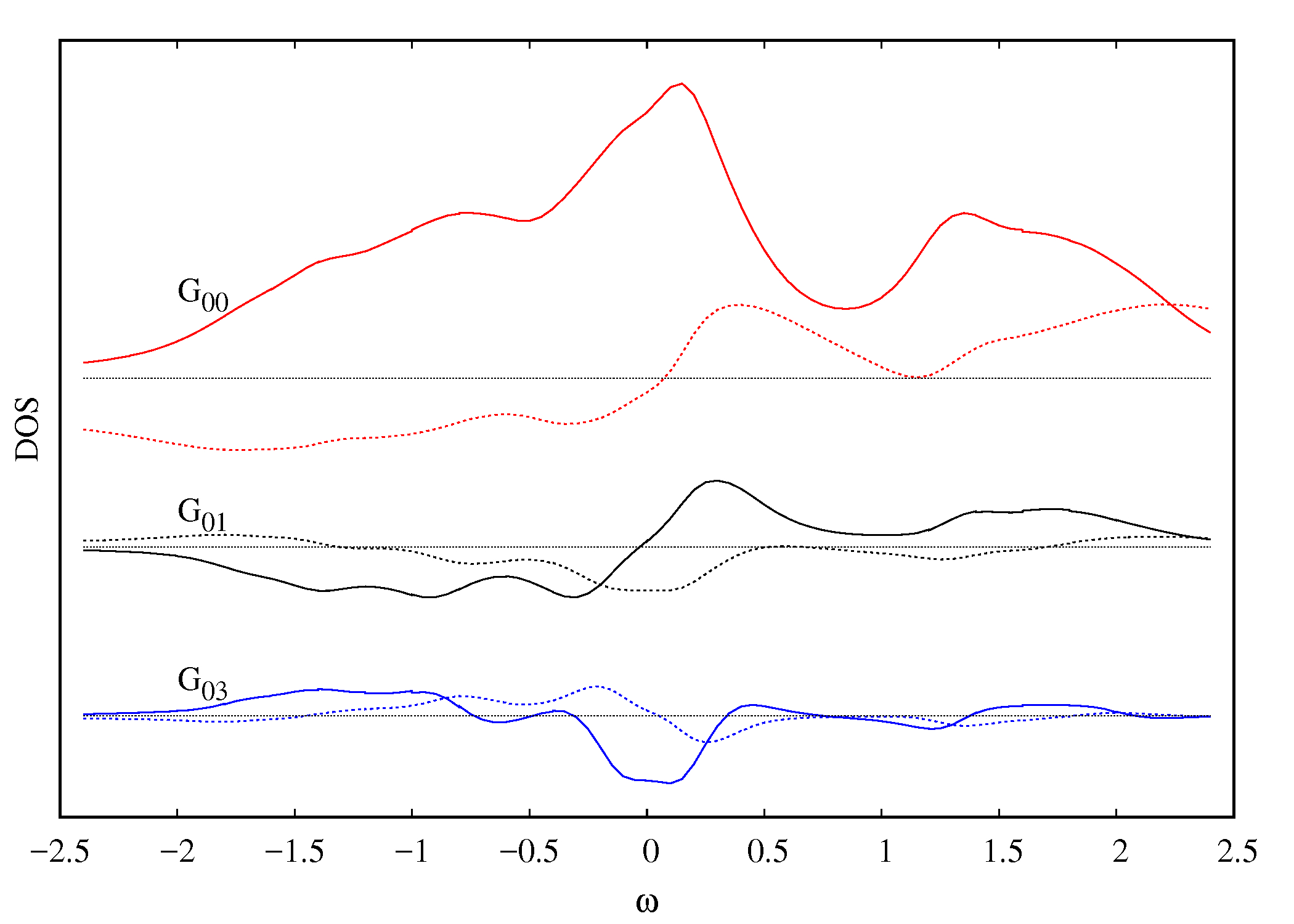}
\par\end{centering}
\caption{\label{fig4} Imaginary (continuous lines) and real (dotted lines) Green's functions for the doped Hubbard model ($\mu=-0.3$) on the square lattice with $t'=-0.05$ and $U=2$ calculated using c4-DMFT, arbitrary units. The red continuous curve corresponds to the density of states. The Fermi energy lies at $\omega=0$ and the horizontal lines are at zero. }
\end{figure}

To illustrate the results with finite doping, in Fig. \ref{fig4} we show the DOS for the Hubbard Hamiltonian on the square lattice with nearest ($t$) and a finite next nearest-neighbor hopping ($t'$), together with the non-local Green's functions.

\section{Conclusions}

We have presented here an efficient and reliable numerical method to calculate dynamical properties of complex impurities based on the DMRG. This technique uses the correction vector to obtain precise Green's functions on the real frequency axis directly thus avoiding ill-posed analytic continuation methods from the Matsubara frequencies and fermionic sign problems present in quantum Monte Carlo-based techniques, allowing also for
zero temperature calculations. When used as the impurity-solver of the DMFT algorithm it leads to highly reliable spectral functions by using a self-consistent bath with low entanglement for which the density matrix renormalization works best. 

To illustrate the versatility of the method, we have shown examples of densities of state and response functions within the DMFT framework for two paradigmatic models such as the Hubbard model at half filling on the square lattice on the one, two and four-site effective impurity models and at finite doping on the four-site case and also for the two-band Kanamori-Hubbard model on the Bethe lattice in the presence of Hund's coupling and interband hybridization. 

This method leads to reliable results for non-local self energies at arbitrary dopings, hybridizations and interactions, at any energy scale. It also paves the way to treating large effective impurities not only within the framework of the DMFT to study multi-band interacting models and multi-site or multi-momenta clusters, but also for complex impurity problems such as adsorbed atoms, cold atoms and interacting nanoscopic systems like quantum dot arrays among others. 

There is room to include additional improvements such as the consideration of symmetries, finite temperature, and more realistic systems by taking into account configurations given by ab-initio methods.

\section*{Acknowledgments}
We acknowledge support from projects PICT 2012-1069 and PICT 2016-0402 from the Argentine ANPCyT and PIP 2015-2017 11220150100538CO (CONICET). This work used the Extreme Science and Engineering Discovery Environment (XSEDE), which is supported by National Science Foundation grant number ACI-1548562 and is also funded in part by a QuantEmX grant from ICAM and the Gordon and Betty Moore Foundation through Grant GBMF5305 to K. Hallberg. We thank G. Kotliar, D. Garc\'{\i}a, P. Cornaglia, M. Imada and S. Sakai for useful discussions.



\bibliographystyle{frontiersinHLTH&FPHY} 
\bibliography{biblio}






\end{document}